\documentclass[fleqn,twoside]{article} 
\usepackage{epsf,multicol,ifthen}
\usepackage{ujp}
\usepackage[cp1251]{inputenc}
\usepackage[english,russian]{babel}
\usepackage{amstext}
\usepackage{amssymb}
\usepackage{cite}
\usepackage {graphicx,amsmath}

\nazva{DETERMINATION OF THE EFFECTIVE PARAMETERS\\ OF~
PROTON\,--\,\boldmath$^{3}$He~
SCATTERING~ ON~ THE~ BASIS\\ OF THE NEUTRON-TRITON SCATTERING DATA}%

\udk{539.171 }

\nazvacol{DETERMINATION OF THE EFFECTIVE PARAMETERS}%

\avtor{V.P. LEVASHEV}%
\avtorcol{V.P. LEVASHEV }%

\inst{M.M. Bogolyubov Institute for Theoretical Physics, Nat. Acad.  Sci. of Ukraine}%
\adr{(14b, Metrolohichna Str., Kyiv 03143, Ukraine; e-mail:
levashev@bitp.kiev.ua) }

\begin{document}           
\setcounter{page}{436}%
\maketitl                 
\begin{multicols}{2}
\anot{%
We have studied the relations between the neutron-triton
scattering lengths and effective ranges and the corresponding
quantities for the p\,--$^{3}$He scattering in the framework of
the potential model with an effective nucleon-nucleus interaction
in the form of a $\delta $-shell potential. It is shown that the
Coulomb renormalization of the pure nuclear scattering lengths
does not change the relation well established for the n + $^{3}$H
system between the lengths: $A^{1} < A^{0}$. We have predicted the
p--$^{3}$He scattering lengths which give preference to set I of
the phase analysis performed by E.A. George et al. (2003), which
corresponds to the inequality
$A^{1}_{nc} < A^{0}_{nc}$ for the scattering lengths.}%

\section{Introduction}

The topicality of the determination of parameters of the
low-energy scattering in few-nucleon systems consists in the
possibility to verify the theoretical approaches and models used
for the description of such systems. These parameters are also
used as the initial data in the calculations of the more
complicated processes with participation of weak forces (for
example, for the weak capture of a proton by light nuclei). At
present, the low-energy scattering parameters for the physical
systems with four nucleons are known with various degrees of
accuracy. In some cases, they are so indeterminate that there
exist, for example, two essentially different sets of scattering
lengths (zero-energy scattering amplitudes). In this connection,
of great interest is the problem to study the correlation
relations between different characteristics of different
four-nucleon systems, which would allow one to obtain the
independent information about insufficiently well-measured
quantities on the basis of reliable available data.

Here, on the basis of the potential model with an effective
nucleon-nucleus interaction in the form of $\delta$-shell
potentials, we study the relation between the lengths and
effective ranges of the $S$-wave scattering of a proton by a
nucleus $^{3}$He (h, in brief) and the corresponding quantities
for the scattering of a neutron by a triton $^{3}$H (t, in brief).

\section{Analysis of the Data on p\boldmath$-^{3}$He Scattering}

The importance of the consideration of this problem is conditioned
by the experimental situation observed now in the determination of
low-energy  parameters of the scattering of a proton by a $^{3}$He
nucleus. In Table 1, we present values of the nuclear-Coulomb
scattering lengths $A^{1}_{nc}$ and $A^{0}_{nc}$, which are,
respectively, a triplet and a singlet by the total spin, and the
corresponding effective ranges $r^{1}_{nc} $ and $  r^{0}_{nc} $
found on the basis of the results of the phase analyses of
experimental data on the p--$^{3}$He scattering [1--6]. The
asterisk marks the results obtained by us on the basis of the data
of the corresponding works.

The most full phase analysis of all the totality of data for the
p--$^{3}$He scattering performed in the recent work [6] has not
given an unambigous set of values for the triplet and singlet
scattering lengths of the p-h scattering. In that work, two sets
of the scattering lengths were presented.
\begin{equation}
A^{1}_{nc}=7.9~\text{fm},\quad A^{0}_{nc} =15.1~\text{fm}\quad
\text{set I};
\end{equation}\vskip-8mm
\begin{equation}
A^{1}_{nc}=10.4~\text{fm},\quad A^{0}_{nc} =7.2~\text{fm}\quad
\text{set II}.
\end{equation}

These sets differ qualitatively and quantitatively from each other. Solutions
(1) and (2) were obtained on the basis of the dataset in [5], to which the new
data on the proton analyzing power [7] and the data on the
cross-sections and the analyzing powers at very low energies
[8]. Just the last data significantly affected the final result.
We note that set I is characterized by a less value of the function $\chi ^{2} =
0.958$ per data point, than set II ($\chi ^{2} = 0.973)$, which can indicate
the higher reliability of the former solution.

It is worth noting that the previous phase analyses [1, 2] have
also led to several solutions. From two results related to
solutions a) and c) in [1], Table III, our table presents solution
c) as such that agrees with the empiric observation $r^{1}_{nc}<
r^{0}_{nc}$. The values corresponding to work [2] are extracted by
us from solution III given in [2].

In [4], all the data available at that time on the p--$^{3}$He
scattering up to 10 MeV (scattering cross-sections, polarizations
of a proton and a target, and the polarization transfer
coefficients) were analyzed in the framework of the model of the
separable potential interaction between a nucleon and a
three-nucleon nucleus.

As for the result of work [3], it was obtained by means of the
fitting of triplet phase shifts with the help of the expansion of
the effective range on the basis of the data on phase shifts [8]
and the results of the very old phase analysis [9] which have
great errors. As a result, the effective range obtained in [3]
seemed to be too small. For this reason, the authors of work [3],
with the help of the experimental value of the triplet
neutron-triton scattering length $A^{1}_{n}= (3.68\pm  0.05)$ fm
[10] and the well-known Jackson--Blatt formula [11], got a
somewhat larger value
\begin{equation}
r^{1}_{nc}= (1.30 \pm 0.18) \, \rm fm.
\end{equation}
Nevertheless, the value of $r^{1}_{nc} $ (3) remains noticeably
less than two other values in Table 1.

In order to make an unambiguous choice of one of the two sets of
p--$^{3}$He scattering  lengths (1), (2), we will involve the
experimental and theoretical data on the charge-symmetric
neutron + triton system into the analysis. In the
approximation of the charge symmetry of nuclear forces, the system
n$+^{3}$H$, $which has no Coulomb interaction between a neutron
and a triton, can be an additional source of information in
studying the characteristics of the nuclear-Coulomb system
p$+^{3}$He.

\section{Analysis of the Data on n\boldmath--$^{3}$H scattering}

The experimental situation for the n--t scattering lengths looks
better at least for the reason that these lengths can be directly
determine with the help of any pair of such physical quantities as
the threshold value of the total cross-section
\begin{equation}
\label{eq1}
\sigma = \pi (A_{n}^{0^{2}} + 3A_{n}^{1^{2}} ),
\end{equation}

\noindent
the coherent length
\begin{equation}
\label{eq2}
A_{c} = {\frac{{1}}{{4}}}(A_{nc}^{0} + 3A_{nc}^{1} ),
\end{equation}

\noindent
and the incoherent length
\begin{equation}
\label{eq3}
A_{i} = {\frac{{\sqrt {3}}} {{4}}}(A_{n}^{1} - A_{n}^{0} )
\end{equation}

\noindent of the n--t scattering.

In our first calculations [12, 13] of the triplet $A^{1}_{n}$ and
singlet $A^{0}_{n}$ pure nuclear n--t scattering lengths, we noted
the inconsistency of the experimental data available at that time
on $\sigma $ and $A_{c}$ (see [13]). From that time, the accuracy
of measurements of these quantities is significantly enhanced, and
the contemporary experimental situation is presented in Table 2.
The available experimental data on the n--t scattering
cross-section $\sigma $ [10] and the coherent scattering length
$A_{c} $ [14, 15] impose, according to relations (4), (5), some
empiric restrictions on the lengths $A^{1}_{n}$ and $A^{0}_{n}$.
In the plane of the variables $A^{0}_{n}$ and $A^{1}_{n}$, these
data correspond to the strips, the intersection region of which
gives the expected experimental values of the n--t scattering
lengths (two last columns of Table~2).

We note that work [15], besides the recommended value $A_{c}=
(3.59 \pm  0.02)$ fm, gives also
\begin{equation}
A_{c,{\rm avr}}= (3.61 \pm  0.02)\, \rm fm.
\end{equation}

The last value is obtained by the authors by means of the
averaging of three available values of $A_{c}$. It turns out
however that the small (0.5 {\%}) addition to $A_{c,{\rm avr}}$
(7) leads to the very close approach of the upper edge of the
strip $A_{c,{\rm avr}}$ to the lower edge of the strip related to
the experimental value of $\sigma $ [10]. In this case, a reliable
experimental determination of the lengths $A^{1}_{n}$ and
$A^{0}_{n}$ on the basis of the data for $\sigma $ and $A_{c,{\rm
avr}}$ is quite problematic.

The last row of Table 2 gives the results [16] obtained on the
basis of the $R$-matrix parametrization of the data

\vskip2mm
\noindent{\footnotesize{\bf%
 T a b l e~ 1. Values of the nuclear-Coulomb triplet
($A^{1}_{nc})$ and singlet (\boldmath$A^{0}_{nc})$ scattering
lengths and the corresponding effective ranges
\boldmath$r^{1}_{nc} $ and \boldmath$r^{0}_{nc}$ obtained from the
data on phase shifts of the p\boldmath$-^{3}$He scattering. The
asterix marks the results obtained in the present work on the
basis of the data of the corresponding phase analyses }\vskip1mm
\tabcolsep4.7pt

\noindent\begin{tabular}{c c c c c c}
 \hline \multicolumn{1}{c}
{\rule{0pt}{9pt}Source} & \multicolumn{1}{|c}{$A_{nc}^1$, fm}&
\multicolumn{1}{|c}{$r_{nc}^1$, fm}&
\multicolumn{1}{|c}{$A_{nc}^0$, fm}&
\multicolumn{1}{|c}{$r_{nc}^0$, fm}
& \multicolumn{1}{|c}{}\\%
\hline%
\rule{0pt}{9pt}[1]&7.89&1.62&7.94&1.96&\\%
$[2]$&&&12.8$^*$&1.7$^*$&\\%
$[3]$&$10.2\pm 1,4$&$1.02 \pm 0.42$&&&\\%
$[4]$&7.6$^*$&1.8$^*$&$16.4^*$&$1.9^*$&\\%
$[5]$&$8.1\pm 0.5$&&$10.8\pm 2.6$&&\\%
$[6]$&$7.9\pm 0.2$&&$15.1\pm 0.8$&&set I\\%
&$10.4\pm 0.4$&&$7.2 \pm 0.8$&&set II\\%
 \hline
\end{tabular}
}

\vskip2mm
\noindent{\footnotesize{\bf%
 T a b l e~ 2. Experimental data on triplet (\boldmath$A^{1}_{n})$,
singlet (\boldmath$A^{0}_{n})$, and coherent (\boldmath$A_{c})$
scattering lengths and the total cross-section ($\sigma $) of n--t
scattering}\vskip1mm \tabcolsep2.2pt

\noindent\begin{tabular}{c c c c c c}
 \hline \multicolumn{1}{c}
{\rule{0pt}{9pt}Source} & \multicolumn{1}{|c}{$\sigma$, сэ}&
\multicolumn{1}{|c}{$A_c$, fm}&
\multicolumn{1}{|c}{$A_n^1$, fm}& \multicolumn{1}{|c}{$A_n^0$, fm}& \multicolumn{1}{|c}{}\\%
\hline%
\rule{0pt}{9pt}$[10]$& 1.70 $\pm $ 0.03& & 3.6 $\pm $ 0.1& 3.91
$\pm $ 0.12&
 \\%
$[14]$& & 3.82 $\pm $ 0.07& 3.70 $\pm $ 0.21& 3.70 $\pm $ 0.62&
 \\%
$[15]$& & 3.59 $\pm $ 0.02& 3.13 $\pm $ 0.11& 4.98 $\pm $ 0.29&
set I \\%
& & & 4.05 $\pm $ 0.09& 2.10 $\pm $ 0.31& set II \\%
$[16]$& & & 3.325$\pm $0.016& 4.453$\pm $0.10&\\%
\hline
\end{tabular}
}

\noindent for p--$^{3}$He scattering with the approximate account
of the Coulomb difference of the p--$^{3}$He and n--$^{3}$H
systems.

The results of theoretical calculations of the parameters of
low-energy n--t scattering for the recent years [17--23] are
presented in Table 3. These calculations are mainly based on
solutions of the four-particle equations which take strictly the
multiparticle dynamics and boundary conditions of the problem into
account. As for the interaction between nucleons, the models
including even the three-particle forces, in addition to
complicated two-particle potentials, have been used in the last
years (see, e.g., the third row of the results from [23] in Table
3). The averaging of the theoretical results for scattering
lengths and effective ranges, which give the values of the
coherent scattering length and total n--t scattering cross-section
(Table 2) closely to the experimental data, yields
\begin{equation}
A^{0}_{n}= (4.0 \pm  0.1)\, {\rm  fm}, \quad r^{0}_{n}= (1.95 \pm
0.05)\, {\rm fm},
\end{equation}
\begin{equation}
A^{1}_{n}= (3.58 \pm  0.05)\, {\rm fm}, \quad r^{1}_{n}= (1.75 \pm
0.05)\, {\rm fm}.
\end{equation}
In this case, for the lengths and the effective ranges (8) and
(9), the following relations hold:
\begin{equation}
A^{1}_{n} / A^{0}_{n} = 0.895  < 1,
\end{equation}
\begin{equation}
r^{1}_{n} / r^{0}_{n}  = 0.897 < 1.
\end{equation}

An additional source of the information on the parameters of
low-energy neutron-triton scattering would be the results of phase
analyses of the experimental data in the energy range up to 10
MeV. However, the results of the old phase analysis of n--t
scattering [24] were obtained on the basis of a restricted
dataset, are characterized large errors, and can be used in order
to get the information on the scattering lengths with caution. For
the singlet spin channel, the phase shifts~~ from~~ [24]~~ are~~
very~~ much~~ underestimated~~ by

\vskip3mm
\noindent{\footnotesize{\bf%
 T a b l e~ 3. Theoretical date on triplet (\boldmath$A^{1}_{n})$ and
singlet (\boldmath$A^{0}_{n})$ scattering lengths and the
corresponding effective ranges (\boldmath$r^{1}_{n} $ and
\boldmath$r^{0}_{n})$ and the coherent scattering length
(\boldmath$A_{c})$ and the total cross-section ($\sigma $) of the
scattering of a neutron by a triton}\vskip1mm \tabcolsep5.1pt

\noindent\begin{tabular}{c c c c c c c}
 \hline \multicolumn{1}{c}
{\rule{0pt}{9pt}Source} & \multicolumn{1}{|c}{$A_n^1$, fm}&
\multicolumn{1}{|c}{$r_n^1$, fm}& \multicolumn{1}{|c}{$A_n^0$,
fm}& \multicolumn{1}{|c}{$r_n^0$, fm}&
\multicolumn{1}{|c}{$A_c$, fm}& \multicolumn{1}{|c}{$\sigma$, fm}\\%
\hline%
\rule{0pt}{9pt}$[17]$& 3.61& & 4.09& & 3.73& 175.4 \\%
$ [18]$& 3.46& & 4.24&& 3.66& 169 \\%
$[19]$& 3.597& & 3.905& & 3.68& 170 \\%
$[20]$& 3.6& & 4.0& & 3.7& 172.4 \\%
$[21]$& 3.80& & 4.32& & 3.93& 194.7 \\%
$[22]$& 3.63& & 4.10& & 3.75& 177.0 \\%
& 3.73& 1.87& 4.13& 2.01& 3.83& 184.7 \\%
$[23]$& 3.76& & 4.31& & 3.90& 191.6 \\%
& 3.79& 1.76& 4.31& 2.08& 3.92& 193.7 \\ %
& 3.53& 1.71& 3.99& 1.95& 3.65& 167.5\\%
 \hline
\end{tabular}
}

\noindent modulus and are characterized by great errors.
Generally, the two-parameter approximation of the singlet and
triplet phase shifts from [24] gives
\begin{equation}
A^{0}_{n}= (2.8 \pm  0.5) \, {\rm fm}, \quad r^{0}_{n}\approx 1.6
\, \rm fm,
\end{equation}\vskip-5mm
\begin{equation}
A^{1}_{n}= (3.4 \pm  0.5)\, {\rm  fm},\quad  r^{1}_{n}\approx 2.3
\, \rm fm.
\end{equation}
The result for the singlet length $A^{0}_{n}$ (12) is
significantly less than all available data for this quantity (see
Tables 1 and 2). The same conclusion concerns the effective range
$r^{0}_{n}$ (12). For the triplet channel, the effective range
$r^{1}_{n}$ (13) looks to be very much overestimated. We note that
values (12) and (13) do not agree with relations (10) and (11).

In the other phase shift analysis [4], the predictions for the
n--t scattering phase shifts were made on the basis of the results
of the phase analysis of p--h data in the framework of the model of
separable intercluster interaction. After the optimization of
fitting parameters of the nuclear p--h potential, the authors of
work [4] calculated the phase shifts of p--h scattering and the
scattering phase shifts for the charge-symmetric n--t system. The
results of work [4] let us to extract the following scattering lengths
and effective ranges:
\begin{equation}
A^{0}_{n}= 5.5\, {\rm fm}, \quad r^{0}_{n}\approx 2.3\, \rm fm,
\end{equation}\vskip-5mm
\begin{equation}
A^{1}_{n}= 3.2\, {\rm fm}, \quad r^{1}_{n}\approx 2.1\, \rm fm.
\end{equation}
But, from our viewpoint, just the phase shifts of n--t scattering
in [4] are not sufficiently exact, because they are a
reflection of the phase shifts of p--h scattering. However, as
compared with the later more exact phase analysis [5], the
singlet phase shifts [4] in the interval of laboratory energies
less than 6 MeV look essentially (by 10$^\circ$) overestimated.
Respectively, lengths (14) and (15) yield the total n--t
scattering cross-section $\sigma  =1.92$ b, which exceeds the
experimental result [10] by 13{\%} (see Table 2). We note that the
effective parameters (14) and (15) satisfy relations (10) and
(11).

In what follows as the input data of our analysis of p--h
scattering in the triplet spin channel, we use the experimental
value of the triplet length of n--t scattering,
\begin{equation}
A^{1}_{n}= (3.6 \pm  0.1)\, \rm fm,
\end{equation}

\noindent which agrees well with the theoretical prediction (9).
For the singlet spin channel, we will consider two following
values of the n--t scattering length:
\begin{equation}
A^{0}_{n}= (4.0 \pm  0.1)\, \rm fm,
\end{equation}

\noindent which represent the experimental value [10] and the
averaged theoretical result (8), and
\begin{equation}
A^{0}_{n}= (4.453 \pm  0.100)\, \rm fm,
\end{equation}
being a result of the $R$-matrix parametrization of the data on
$p-^{ 3}$He scattering [16]. The experimental set~I [15] includes
the triplet and singlet lengths of n--t scattering which lie
outside values (16)--(18).

\section{Determination of Effective Parameters of the
p--h Scattering from the n--t Data}

Under assumption of the charge symmetry of nuclear forces, the
available information on the n+t system can be used in the
derivation of the scattering lengths and effective ranges for the
p+h system. In order to characterize the effective nuclear
interaction of a nucleon with a three-nucleon nucleus, we will use
the model of $\delta$-shell potential
\begin{equation}
\label{eq4}
V(r) = - \lambda {\frac{{\delta (r - R)}}{{R^{2}}}},
\end{equation}

\noindent
where $R$ is the interaction radius, and $\lambda$ is the potential intensity.

We note that a potential of the form (19) well describes the
$S$-wave phase shifts of \textit{NN} scattering up to energies
$\simeq $250 $\div $ 300 MeV in the laboratory system and ensures
a good description of low-energy characteristics of a system of
three nucleons [25]. This potential agrees naturally with the fact
that low-energy properties of a three-nucleon system turn out
insensitive to a detailed behavior of the interaction at small
distances and are mainly determined by its intensity at distances
about (1.5 $\div $ 2) fm.

Potential (19) was successfully used earlier as \textit{NN}-forces
in the calculation of the characteristics of threshold n--t
scattering [18, 26, 27] on the basis of the strict four-particle
Faddeev--Yakubovsky integral equations.

For potential (19), the pure nuclear and modified
nuclear-Coulomb scattering lengths and effective ranges are
expressed through the potential parameters $R $ and $\lambda $ in
the analytic form [28]:
\begin{equation}
\label{eq5}
{\frac{{1}}{{A_{n}}} } = {\frac{{1}}{{R}}} - {\frac{{1}}{{\lambda}} },
\end{equation}
\begin{equation}
\label{eq6}
r_{n} = {\frac{{2R^{2}}}{{3}}}({\frac{{1}}{{R}}} + {\frac{{1}}{{\lambda
}}}),
\end{equation}
\begin{equation}
\label{eq7} {\frac{{1}}{{A_{nc}}} } = {\frac{{2}}{{a_{\rm B}
I_{1}^{2}}} }{\left[ { - {\frac{{R}}{{\lambda}} } + 2I_{1} K_{1}}
\right]},
\end{equation}
\begin{equation}
\label{eq8} r_{nc} = {\frac{{2a_{\rm B}}} {{3I_{1}^{2}}} }{\left[
{{\frac{{R}}{{\lambda }}}{\frac{{x^{3 / 2}I_{2}}} {{I_{1}}} } +
{\frac{{1}}{{2}}}(I_{1}^{2} - x)} \right]}.
\end{equation}

In formulas (22) and (23), $a_{\rm B} = \hbar ^{2} / (2\mu e^{2})$
is the Bohr radius of the p--h system, $\mu$ is the reduced mass,
$x = 2R / a_{\rm B} $, and $I_{1}, I_{2}$, and $ K_{1}$ are the
modified Bessel function of the variable $z = 2\sqrt {x} $. As
$\lambda \to \infty $, formulas (20)--(23) take the form
\begin{equation}
\label{eq9}
A_{n}^{\infty}  = R,
\end{equation}
\begin{equation}
\label{eq10}
r_{n}^{\infty}  = {\frac{{2}}{{3}}}R,
\end{equation}
\begin{equation}
\label{eq11} A_{nc}^{\infty}  = {\frac{{a_{\rm B}}}
{{4}}}{\frac{{I_{1} (z)}}{{K_{1} (z)}}},
\end{equation}
\begin{equation}
\label{eq12} r_{nc}^{\infty}  = {\frac{{a_{\rm B}}} {{3}}}(1 -
{\frac{{z^{2}}}{{4I_{1}^{2} (z)}}}),
\end{equation}

\noindent where $z = (8R / a_{\rm B} )^{1 / 2}$.

Below, we will find the potential of the pure nuclear interaction
of a nucleon with a three-nucleon nucleus (parameters $R $ and
$\lambda $) from formulas (20) and (21), where we will use, as the
input values, the most reliable data for the scattering lengths
and effective ranges of a neutron by a triton, $A_{n} $ and $r_{n}
$. Then the nuclear potential constructed in such a way will be
applied to the determination of the scattering lengths and
effective ranges $A_{nc}$ and $ r_{nc}$ for the nuclear-Coulomb
scattering of a proton by a nucleus $^{3}$He.

In order to study the sensitivity of the predictions to the value
of the uncertainty of the n--t data, we use the set of $\delta
$-shell potentials (19), by varying the range of the potential $R
$ and its intensity $\lambda $ in wide limits at fixed values of
the n--t scattering lengths $A_{n}$ or the effective range
$r_{n}$. Such a set includes both the attractive and repulsive
potentials. Within such an approach, we will clarify the character
of the correlation relations between various quantities and use
them for the derivation of an important additional information for
different physical systems. We note that the use of repulsive
potentials can be substantiated by the presence of a strong Pauli
repulsion in the interaction of a nucleon with a three-nucleon
nucleus in the spin states under consideration. Such a repulsion~
counteracts~ the~ localization~ of~ the system~ in~ a~ small~
region~ of~ the~ configuration~ space,

\begin{center} \noindent \epsfxsize=0.95\columnwidth\epsffile{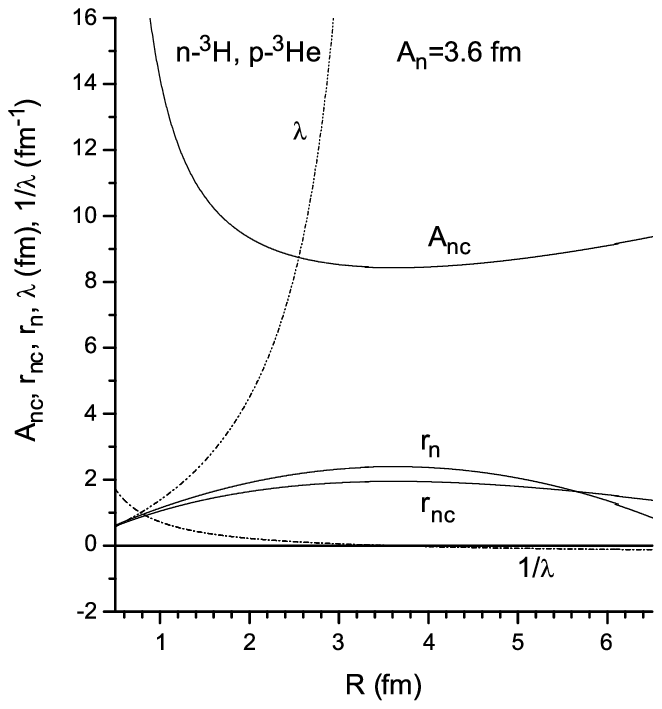}
\end{center}

\vskip-3mm\noindent{\footnotesize Fig. 1. Effective parameters of
triplet p--h scattering $A^{1}_{nc}$ and $ r^{1}_{nc} $, the
range of n--t scattering $r^{1}_{n}$, and the
$\delta$-shell potential intensity $\lambda $ versus the effective
range of the potential $R$ at a fixed experimental value of the
n--t scattering length $A^{1}_{n}=3.60$ fm}%
\vskip15pt

\noindent where the nuclear attraction between nucleons is
significant.

For the the nucleon~+ $\alpha $-particle system, which is
very close to our case as to the character of the intercluster interaction,
such $S$-wave
potential of the repulsive form was successfully used in the
description of the phase shifts of N--$\alpha$ scattering up to
energies $\simeq$ 25 MeV [29]. On the other hand, such systems are
traditionally described with potentials of the attractive type
[30]. The attractive character of the interaction of a nucleon
with an $\alpha $-particle is also demonstrated by results of the
solution of the inverse problem of N--$\alpha$ scattering [31].

In Fig. 1, we present the dependence of the effective parameters
of p--$^{3}$He scattering ($A_{nc}$ and $ r_{nc})$ calculated by
us, the effective range of n--t scattering ($r_{n})$, and the
potential intensity $\lambda $ on the potential range at a fixed
experimental value of the n--t scattering length in the triplet
spin channel $A^{1}_{n}=3.6$ fm [10]. The parameters $R$ and
$\lambda $ of the set of attractive ($\lambda  > 0$) and repulsive
($\lambda < 0$) $\delta $-shell potentials (19) of the nuclear
interaction of a nucleon with a three-nucleon nucleus are varied
in a correlated manner so that to ensure the production of a
chosen fixed four-nucleon parameter $A^{1}_{n}$ [10] with the use of
these potentials.

The value of the radius $R_{\infty} = 3.60$ fm, at which the
function $1 / \lambda $ crosses the abscissa, corresponds to
the scattering by a sphere with the infinite attraction (on the
approach to the point from the left) or by the infinite repulsion
(on the approach from the right). The last variant is known in the
literature as the hard-sphere model which was quite successfully
used on the early stages of the studies of the processes of
scattering in channels containing no bound or resonant states of
the system. At the point $R_{\infty}$, the plot of the function
$A_{nc}(R)$ reaches a minimum, $A^{\infty} _{nc}= 8.4$ fm, whereas
the functions of effective ranges take the maximum values
$r^{1}_{nc}=1.95$ fm and $ r^{1}_{n}=2.40$ fm at this point. The
presented numerical values can be obtained directly with the help
of formulas (24)--(27) which imply also that the value of
$R_{\infty} $ coincides with the length $A^{\infty} _{n} $ (24).
To the left and to the right from the point $R_{\infty}$, values
of $A_{nc}$, $r_{n}$, and $ r_{nc} $ which correspond to,
respectively, attractive and repulsive potentials are positioned.
We note that the relation
\begin{equation}
r_{n} > r_{nc}
\end{equation}
holds for all attractive interactions. For repulsive potentials,
inequality (28) at $R \approx 5.63$ fm (for the given  specific
plot) changes into the inverse one:
\begin{equation}
r_{n} < r_{nc}.
\end{equation}

Figure 1 demonstrates clearly how, for example, the choice of a specific value
of the effective range $r_{n}$ of n--t scattering
allows one to at once determine a value of $R $ (and, hence,
$\lambda (R))$ and to obtain the nuclear-Coulomb quantities
$A_{nc}(R)$ and $ r_{nc}(R)$ corresponding to these particular pure nuclear
quantities $A_{n} $ and $r_{n}$. Analogously,
as the second quantity (in addition to $A_{n}=3.6$ fm) for the
fixation of specific parameters of the potential, $R $ and
$\lambda(R)$, we can choose $A_{nc}$ or $ r_{nc} $ and to find,
respectively, $r_{n} $ and $r_{nc} $ or $A_{nc}$ and $r_{n}$.

In Fig. 2, we present the correlations between the effective
parameters of triplet p--h scattering, $A^{1}_{nc}$ and
$r^{1}_{nc}$, and the effective range of triplet n--t scattering
$r^{1}_{n}$ at a fixed experimental value of the triplet length of
n--t scattering, $A^{1}_{n} =3.6$ fm [10]. These interrelations
directly between four-nucleon characteristics were obtained by the
internal parametrization of the calculated quantities in terms of
the range of the potential, $R$. We note that the
analytic formulas for the correlation relations shown in the plots
follow from (17)--(20):
\begin{equation}
\label{eq13} {\frac{{1}}{{A_{nc}}} } = {\frac{{2}}{{a_{\rm B}
I_{1}^{2}}} }{\left[ {{\frac{{R}}{{A_{n}}} } + 2I_{1} K_{1} - 1}
\right]},
\end{equation}
\begin{equation}
\label{eq14} {\frac{{1}}{{A_{nc}}} } = {\frac{{2}}{{a_{\rm B}
I_{1}^{2}}} }{\left[ {1 + 2I_{1} K_{1} - {\frac{{3}}{{2R}}}r_{n}}
\right]},
\end{equation}
\begin{equation}
\label{eq15}
{\frac{{1}}{{A_{nc}}} } = f_{1} (R) + f_{2} (R)r_{nc} ,
\end{equation}
\begin{equation}
\label{eq16} f_{1} (R) = {\frac{{1}}{{a_{\rm B}}} }{\left[
{4{\frac{{K_{1}}} {{I_{1}}} } + {\frac{{I_{1}^{2} - x}}{{x^{3 /
2}I_{1} I_{2}}} }} \right]}, \quad f_{2} (R) = {\frac{{3I_{1}}}
{{a_{\rm B}^{2} x^{3 / 2}I_{2}}} },
\end{equation}
\begin{equation}
\label{eq17} r_{nc} = {\frac{{2a_{\rm B}}} {{3I_{1}^{2}}} }{\left[
{{\frac{{I_{2}}} {{I_{1} }}}x^{3 / 2}({\frac{{3}}{{2R}}}r_{n} - 1)
+ {\frac{{1}}{{2}}}(I_{1}^{2} - x)} \right]}.
\end{equation}
In fact, the graphical interpretation of relations (30)--(34)
allows us to obtain a numerous additional information in addition
to the numerical one.

In Fig. 2, we give also the values of the range $R$ and the
intensity $\lambda$ of potentials (\ref{eq4}) which reproduce the
above-mentioned experimental value of the nuclear scattering
length $A^{1}_{n}$, but give different values of the effective
range $r^{1}_{n}$. The extreme right points of the plots in Fig. 2
correspond to the potentials with $\lambda =\pm \infty $ and
denote the transition from the set of attractive potentials
(branches (+)) to the set of repulsive potentials (branches (--)).
The values of $A_{nc}, r_{nc} $ and $r_{n}$ obtained from this
plot as ${\left| {\lambda}  \right|} = \infty $ reconstruct
exactly the numerical values given above directly prior to formula
(28).

The analysis of the character of the dependence of the quantities
$A_{nc}$ and $ r_{nc}$ on $r_{n}$ indicates that the triplet
p--$^{3}$He scattering length for repulsive potentials decreases
very weakly in a linear manner from 9.1 to 8.4 fm ($ \simeq $8
{\%}) under a quite significant variation of the effective range
of n--t scattering $r_{n} $ in the scope 1.5 $\div $ 2.4 fm (=60
{\%}). In this case, the nuclear-Coulomb effective range $r_{nc}$
increases by $ \simeq $23 {\%} from 1.6 to 1.95~fm.

For the attractive potentials (19), the nuclear-Coulomb quantities
$A_{nc}$ and $ r_{nc }$ show a considerably greater sensitivity to
the variation of the effective range of n--t scattering. With
increase in $r_{n}$ from 1.5 to 2.4 fm, the length $A_{nc}$
decreases by $ \simeq $23 {\%} from 10.97 to 8.43 fm, and the
effective range $r_{nc}$ demonstrates a twice greater sensitivity,
by growing by 46 {\%} from 1.34 to 1.95~fm.

The above-presented results allow us to conclude that

1) scattering lengths $A_{nc}$ are less (by 2--3 times) sensitive
to variations of the effective range of n--t scattering $r_{n} $,
than the nuclear-Coulomb effective range $r_{nc}$;

\begin{center} \noindent \epsfxsize=\columnwidth\epsffile{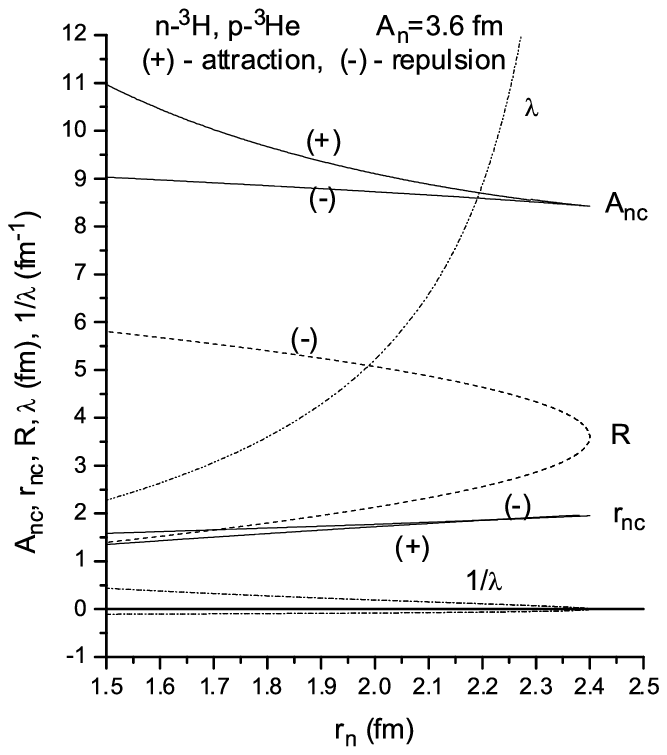}
\end{center}

\vskip-3mm\noindent{\footnotesize Fig. 2. Correlations between the
effective parameters of triplet p--h scattering $A^{1}_{nc}$ and
$r^{1}_{nc}$ and the effective range of triplet n--t scattering
$r^{1}_{n}$ at a fixed experimental value of the triplet n--t
scattering length $A^{1}_{n} =3.6$ fm [10]. The values of the
range of the $\delta$-shell potential $R$ and the
intensity $\lambda$ ((+) -- attractive, (--) -- repulsive), which
reproduce
the low-energy parameters presented in the plot, are also given}%
\vskip15pt

\noindent

 2) in the case of attractive potentials, the sensitivity
of the quantities $A_{nc}$ and $ r_{nc}$ to the variation of the
effective range of n--t scattering $r_{n}$ is greater,
respectively, by $ \simeq $3 and $ \simeq $2 times, than that for
repulsive forces.

In Figs. 3 and 4, we give the correlations between the effective
parameters of the singlet p--h scattering $A^{0}_{nc}$ and
$r^{0}_{nc}$ and the effective range of the singlet n--t
scattering $r^{0}_{n}$ at fixed values $A^{0}_{n}=4.0$ fm (19) and
$A^{0}_{n}=4.453$ fm [16] given by the $R$-matrix processing of
experimental p--h  data. On the whole, the plots give the same
pattern as that in Fig. 2. The larger nuclear singlet n--t lengths
have led only to a relative increase in the nuclear-Coulomb
effective parameters $A^{0}_{nc}$ and $r^{0}_{nc}$.

In the frame of the developed approach to $\delta $-shell
potentials of the repulsive type, we make prediction for the
effective parameters of p--h scattering on the basis of the data
for the n--t system. For example, for the triplet scattering, the
experimental value~ $A^{1}_{n} =(3.6 \pm  0.1)$ fm

\begin{center} \noindent \epsfxsize=\columnwidth\epsffile{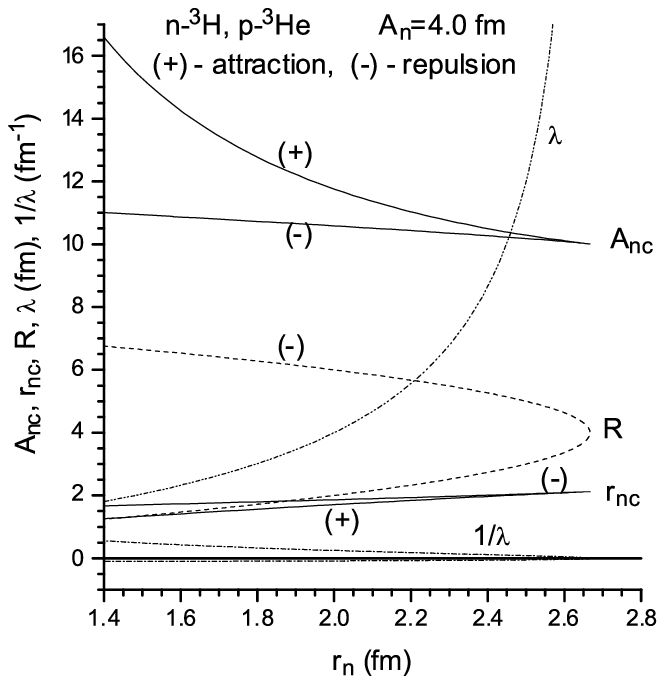}
\end{center}

\vskip-3mm\noindent{\footnotesize Fig. 3. Correlations between the
effective parameters of singlet p--h scattering $A^{0}_{nc}$ and $
r^{0}_{nc} $ and the effective range of singlet n--t scattering
$r^{0}_{n}$ at a fixed experimental value of the singlet n--t
scattering length $A^{0}_{n} =4.0$ fm (17) (Other designations are
the same as in Fig. 2)}%
\vskip15pt

\noindent [10] and the theoretically calculated value
$r^{1}_{n}=(1.75 \pm  0.05$) fm (9) let us to obtain the following
values of the p--h parameters:
\begin{equation}
A^{1}_{nc}= (8.88 \pm  0.48)\, {\rm fm},\quad r^{1}_{nc}= (1.68
\pm
 0.04)\, \rm fm.
\end{equation}
For the scattering in the singlet spin channel, the experimental
and theoretical n--t data on $A^{0}_{n}= (4.0 \pm  0.1)$ fm (17)
and $r^{0}_{n}= (1.95 \pm  0.05)$ fm (8) in Fig. 3 correspond to
the p--h effective parameters
\begin{equation}
A^{0}_{nc}= (10.63 \pm  0.52)\, {\rm fm},\quad r^{0}_{nc}= (1.85
\pm 0.04)\, \rm fm.
\end{equation}
Analogously, for the $R$-matrix result $A^{0}_{n}= (4.453 \pm
0.100)$ fm [16] and $r^{0}_{n}= (1.95 \pm  0.05)$ fm (8), we get
(see Fig. 4)
\begin{equation}
A^{0}_{nc}= (13.05 \pm  0.62)\, {\rm fm},\quad r^{0}_{nc}= (1.97
\pm 0.04)\, \rm fm.
\end{equation}
 Result (37) is somewhat overestimated due to the 11{\%}
overestimation of the input value $A^{0}_{n}=4.453$ fm as compared
with $A^{0}_{n}= 4.0$ fm (8).

For the sake of completeness, we present also values of the
nuclear-Coulomb scattering lengths and effective ranges which
correspond~ to~ attractive~ potentials.~ For

\begin{center} \noindent \epsfxsize=\columnwidth\epsffile{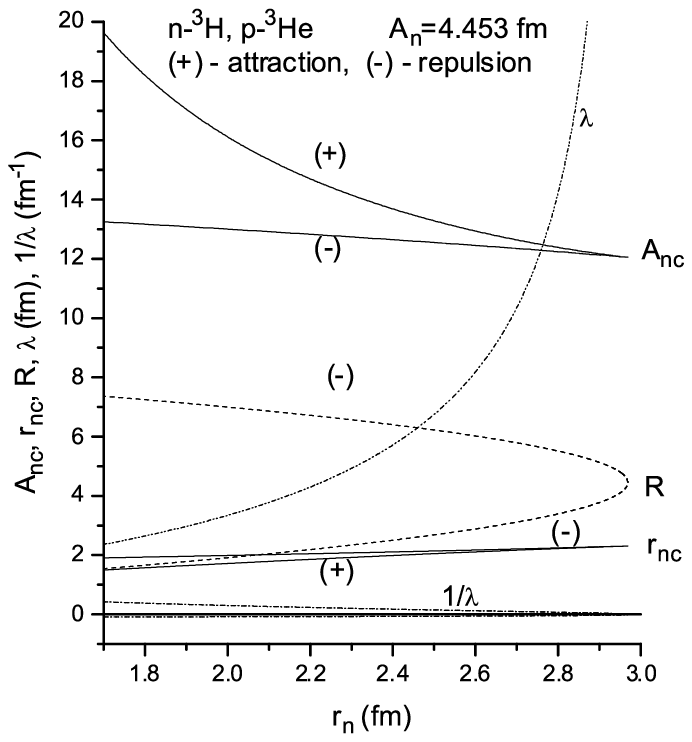}
\end{center}

\vskip-3mm\noindent{\footnotesize Fig. 4. Correlations between the
effective parameters of singlet p--h scattering $A^{0}_{nc}$ and $
r^{0}_{nc} $ and the effective range of singlet n--t scattering
$r^{0}_{n}$ at a fixed value of the singlet n--t scattering length
$A^{0}_{n} =4.453$ fm which corresponds to the $R$-matrix
processing of the experimental p--h data [16] (Other designations
are
the same as in Fig. 2)}%
\vskip15pt

\noindent example, the triplet n--t data, $A^{1}_{n} = (3.6 \pm
0.1)$ fm and $r^{1}_{n}= (1.75 \pm 0.05)$ fm, in Fig. 2 (branch
(+)) correspond to the p--h effective parameters
\begin{equation}
A^{1}_{nc}= (9.90 \pm  0.85)\, {\rm  fm}, \quad r^{1}_{nc}= (1.53
\pm  0.04)\, \rm  fm.
\end{equation}
Analogously, for the singlet n--t data (8), we get (see Fig.~3)
\begin{equation}
A^{0}_{nc}= (12.0 \pm  1.0)\, {\rm  fm}, \quad r^{0}_{nc}= (1.68
\pm
 0.04)\, \rm  fm,
\end{equation}
\noindent whereas, for $A^{0}_{n}= (4.453 \pm  0.100)$ fm and
$r^{0}_{n}= (1.95 \pm  0.05)$ fm, we have (Fig. 4)
\begin{equation}
A^{0}_{nc}= (16.7 \pm  1.7)\, {\rm fm}, \quad r^{0}_{nc}= (1.69
\pm
 0.05)\, \rm fm.
\end{equation}
Result (40) for $A^{0}_{nc}$ is somewhat overestimated because, on
its determination, we used the value of the effective n--t range
$r^{0}_{n}$ (8) which is consistent with the calculated value of
the n--t length near (4.0 $\pm $ 0.1) fm. But, with increase in
the n--t scattering length $A_{n}$, the corresponding effective
range$ r_{n}$ is also increased, as a rule. Therefore, we may
assume that the scattering length $A^{0}_{n}=4.453$ fm must
correspond to the greater effective range $r^{0}_{n}  \simeq $2.15
fm. In this case, the correlation dependence
$A^{0}_{nc}(r^{0}_{n})$ gives a less value, $A^{0}_{nc} \simeq
15.0 $ fm.

Attractive potentials give the values of scattering lengths which
exceed our predictions (35)--(37) for the repulsive effective
interaction by 12{\%}. On the other hand, the effective ranges for
repulsive potentials exceed the corresponding values for
attractive potentials by~10{\%}.

The obtained triplet p--$^{3}$He scattering length (35) well
agrees with the four-particle calculations of this quantity on the
basis of the Kohn variational principle and the technique of
hyperspheric harmonics with regard for three-particle forces [21]:
\[A^{1}_{nc} = 9.13\, \rm fm~ (potential~ AV18UR).\]
A somewhat greater value is obtained in work [21] for potential
AV18 without regard for three-nucleon forces:
\[A^{1}_{nc} = 10.1\, \rm fm.\]
The same value is obtained  in calculations in the frame of the
Monte Carlo method with inclusion of three-particle forces
(potential AV18UR) [32]:
\[A^{1}_{nc} = (10.1 \pm  0.5)\, \rm  fm.\]
Our prediction for the singlet p--h scattering length (36) is,
again, rather close to the result of calculations in [21] for the
interaction including three-particle forces:
\[A^{0}_{nc} = 11.5\, \rm fm~ (potential~ AV18UR).\]
Without regard for three-particle forces, a somewhat greater value
\[A^{0}_{nc} = 12.9\, \rm fm ~(potential~ AV18)\]
was obtained in [21].

The efficiency of the applied method of the
study of the scattering of a proton and a neutron in
charge-symmetric four-nucleon systems can be also demonstrated by
the example of the scattering of a proton and a neutron by an
$\alpha $-particle. The experimental data for scattering lengths
and effective ranges for these systems are known with a quite high
accuracy [33]. Using the data on the scattering of a proton by an
$\alpha $-particle, we determined the attractive $\delta
$-potential and the repulsive potential which is characterized by
a twice greater effective range and the modulus of the coupling
constant (Table 4). The calculations performed with these
potentials gave the convergent values of the scattering length and
the effective range of the scattering in the $n+\alpha$-particle
system. The pure nuclear scattering length obtained from the
nuclear-Coulomb experimental data coincides with the experimental
values obtained from the phase shifts of the scattering of a
neutron by an $\alpha$-particle.  The effective range of the
pure nuclear scattering predicted by us is also very close to
the experimental result for the n+$\alpha$ system. Thus, the model
of $\delta $-shell interaction allows us to get the well
consistent description of the nuclear-Coulomb and nuclear
charge-symmetric systems p+$\alpha $ and n+$\alpha$.

\section{Conclusions}

On the basis of the sufficiently reliable data on the n--t
scattering lengths and the obtained correlation relations for
repulsive potentials, we have predicted the scattering lengths and
effective ranges for the nuclear-Coulomb system p$+^{3}$He:
$A^{1}_{nc}= (8.88 \pm  0.48)$ fm, $r^{1}_{nc} = (1.68 \pm  0.04)$
fm and $A^{0}_{nc}= (10.63\pm 0.52)$ fm, $r^{0}_{nc}= (1.85 \pm
0.04)$~fm.

It is shown that the Coulomb renormalization of pure nuclear
lengths does not change the well-established relation $A^{1}$ <
$A^{0}$ for the n+t system.

We have established that, on the description of the effective
nuclear interaction of a nucleon with a three-nucleon nucleus with
repulsive potentials, the nuclear-Coulomb scattering length and
the effective range of p--h scattering depend linearly on the
effective range of n--t scattering, $r_{n}$. For attractive
potentials, the relevant dependences differ somewhat from linear
ones.

On the whole, our predictions for the p--$^{3}$He scattering
lengths give preference to the results of the phase-shift analysis
in [6] corresponding to the inequality $A^{1}_{nc} <  A^{0}_{nc} $
(set I). Our consideration shows, however, that~ the~ singlet~
scattering~ length~ in~ [6]~ is~ somewhat

\vskip3mm
\noindent{\footnotesize{\bf%
 T a b l e~ 4. Scattering length \boldmath$A_{n}$, the effective range
\boldmath$r_{n}$, and the shape parameter \boldmath$P $ of the
scattering of a neutron by an \boldmath$\alpha $-particle
calculated on the basis of the experimental data for the
nuclear-Coulomb \boldmath$S$-wave scattering of a proton by an
$\alpha $-particle: \boldmath$A_{nc}= (4.97 \pm  0.12)$ fm,
\boldmath$r_{nc}= (1.30 \pm  0.08)$ fm [33] ($\lambda $ is the
intensity, and \boldmath$R$ is the range of the $\delta
$-shell potential of the interaction of a nucleon and an $\alpha
$-particle)}\vskip1mm \tabcolsep2.0pt

\noindent\begin{tabular}{c c c c c c}
 \hline \multicolumn{1}{c}
{\rule{0pt}{9pt}$r_{nc}$, fm} & \multicolumn{1}{|c}{$\lambda$,
fm}& \multicolumn{1}{|c}{$R$, fm}&
\multicolumn{1}{|c}{$A_n$, fm}& \multicolumn{1}{|c}{$r_n$, fm}& \multicolumn{1}{|c}{$P$}\\%
\hline%
\rule{0pt}{9pt}1.38& 9.8259& 1.9895& 2.494& 1.5948& $-0.0693$ \\%
& $-13.8474$& 3.0471& 2.4975& 1.5844& $-0.0537$ \\%
1.30& 5.3867& 1.6919& 2.4667& 1.4822& $-0.0642$ \\%
& $-9.4010$& 3.3667& 2.4789& 1.4407& 0.0211 \\%
1.22& 3.7785& 1.4812& 2.4362& 1.3746& $-0.0605$ \\%
& $-7.7854$& 3.5996& 2.4615& 1.2902& 0.1868 \\%
Exp. $[33]$& & & 2.4641$\pm $0.0037& 1.385$\pm $0.041& \\%
\hline
\end{tabular}
}

\noindent overestimated (by $ \simeq $ 35{\%}) and that the
Coulomb renormalization cannot result in such an increase in the
singlet n--t scattering length. We note that the phase analyses
performed prior to work [6] also gave not quite reliable results for
the singlet phase shift of p--h scattering. In view of the above
discussion, we consider the set
[6]
\[A^{1}_{nc}= (8.2 \pm  0.6)\, {\rm fm},\quad A^{0}_{nc}= (10.3\pm  2.7) \, \rm
fm\]

\noindent as the most reliable phase shift prediction of the p--h scattering
lengths, which corresponds to the p--h dataset of [5] supplemented
by the experimental values of the proton analyzing power [7].

The performed study indicates the necessity to carry out a new
phase analysis of the data on the low-energy p--$^{3}$He
scattering with the use of the scattering lengths predicted by us
and the new experimental data for the differential cross-sections
and the proton analyzing powers obtained in [34] as the input
data.

\end{multicols}

\begin{thebibliography}{99}

\bibitem{1}L. Drigo and G. Pisent, Nuovo Cimento \textbf{51B}, 419 (1967).

\bibitem{2} L.W. Morrow and H. Haeberli, Nucl. Phys. A\textbf{126}, 225
(1969).

\bibitem{3} P.E. Tegner and C. Bargholtz, Astrophys. J. \textbf{272}, 311
(1983).

\bibitem{4} L. Beltramin, R. del~Frate, and G.L. Pisent,~Nucl. Phys.
\textbf{A442}, 266 (1985).

\bibitem{5} M.T. Alley and L.D. Knutson, Phys. Rev. C\textbf{48}, 1901
(1993).

\bibitem{6} E.A. George and L.D. Knutson, Phys. Rev. C\textbf{67}, 027001
(2003).

\bibitem{7} M. Viviani, A. Kievsky, S. Rosati, E.A. George, and L.D.
Knutson, Phys. Rev. Lett. \textbf{86}, 3739 (2001).

\bibitem{8} H. Berg, W. Arnold, E. Huttel, H.H. Krause, J. Ulbricht, and G.
Clausnitzer, Nucl. Phys. A\textbf{334}, 21 (1980).

\bibitem{9} T.A. Tombrello, Phys. Rev. \textbf{138}, B40 (1965).

\bibitem{10} J.D. Seagrave , B.L. Berman, and T.W. Phillips, Phys. Lett.
\textbf{91B}, 200 (1980).

\bibitem{11} J.D. Jackson and J.M. Blatt, Rev. Mod. Phys. \textbf{22}, 77
(1950).

\bibitem{12} V.F. Kharchenko and V.P. Levashev, Phys. Lett. B \textbf{60},
317 (1976).

\bibitem{13} V.F. Kharchenko and V.P. Levashev, Nucl. Phys. A\textbf{343},
249 (1980).

\bibitem{14} S. Hammerschmied, H. Rauch, H. Clerc, and U. Kischko, Z. Phys.
A\textbf{302}, 323 (1981).

\bibitem{15} H. Rauch, D. Tuppinger, H. Wolwitsch, and . Wroblewski, Phys.
Lett. \textbf{165B}, 39 (1985).

\bibitem{16} G.M. Hale et al., Phys. Rev. C\textbf{42}, 438 (1990).

\bibitem{17} J. A. Tjon, Phys. Lett. \textbf{63B}, 391 (1976).

\bibitem{18} V.P. Levashev, Yad. Fiz. \textbf{38}, 566 (1983).

\bibitem{19} A.C. Fonseca, Few-Body Systems \textbf{1}, 69 (1986).

\bibitem{20} S.L. Yakovlev and I.N. Filikhin, Yad. Fiz. \textbf{60}, 1962
(1997).

\bibitem{21} M. Viviani, S. Rosati, and A. Kievsky, Phys. Rev. Lett.
\textbf{81}, 1580 (1998).

\bibitem{22} F. Ciesielski and J. Carbonell, Phys. Rev. C\textbf{58}, 58
(1998).

\bibitem{23} F. Ciesielski, J. Carbonell, and C. Gignoux, Phys. Lett.
B\textbf{447}, 199 (1999).

\bibitem{24} T.A. Tombrello, Phys. Rev. \textbf{143}, 772 (1966).

\bibitem{25} R. van Wageningen, in {\it Proc. of Symp. on Present Status and
Novel Developments in Many{\-}Body Problem}, Rome, September,
1972.

\bibitem{26} V.P. Levashev, Preprint ITP-82-170R, Kyiv, 1983.

\bibitem{27} V.P. Levashev, Preprint ITP-92-65E, Kyiv, 1992.

\bibitem{28} L.P. Kok, J.W. de Maag, H.H. Brouwer, and H. van Haeringen,
Phys. Rev. C\textbf{26}, 2381 (1982).

\bibitem{29} B.V. Danilin, M.V. Zhukov, S.N. Ershov, F.A. Gareev, R.S.
Karmanov, I.S. Vaagen, and J.M. Bang, Phys. Rev. C\textbf{43},
2835 (1991).

\bibitem{30} S. Ali, A.A.Z. Ahmad, and N. Ferdous, Rev. Mod. Phys.
\textbf{57}, 923, (1985).

\bibitem{31} N. Alexander, K. Amos, B. Apagyi, and D.R. Lun, Phys. Rev. C
\textbf{53}, 88 (1996).

\bibitem{32} J. Carlson, D.O. Riska, R. Schiavilla, and R.B. Wiringa, Phys.
Rev. \textbf{C44}, 619 (1991).

\bibitem{33} R.A. Arndt, D.D. Long, and L.D. Roper, Nucl. Phys A\textbf{209},
429 (1973).

\bibitem{34} B.M. Fisher, C.R. Brune, H.J. Karwowski, D.S. Leonard, E.J.
Ludwig, T.C. Black, M. Viviani, A. Kievsky, and S. Rosati,
arXiv:nucl-ex/0608024 v2 (2006).
\begin{flushright}
{\footnotesize Received 13.11.06}
\end{flushright}
\end{thebibliography}
\end{document}